# *Towards an Open Science definition as a political and legal framework: on the sharing and dissemination of research outputs*

**Teresa Gomez-Diaz**
Laboratoire d'informatique Gaspard-Monge (LIGM),
Centre National de la Recherche Scientifique (CNRS), University of Gustave Eiffel (UGE),
Marne-la-Vallée, France
Teresa.Gomez-Diaz@univ-eiffel.fr, orcid.org/0000-0002-7834-145X

**Tomas Recio**
University of Nebrija, Madrid, Spain
trecio@nebrija.es, orcid.org/0000-0002-1011-295X

**V3, 28th February 2021**

**Versions of this document**

| Version | Title | Date | Publication/Dissemination |
|---|---|---|---|
| **V3** | *Towards an Open Science definition as a political and legal framework: on the sharing and dissemination of research outputs* | 28/02/2021 | arXiv, HAL, POLIS, Zenodo. Includes reference: Alma Swan, UNESCO, 2012 |
| **V2** | *Towards an Open Science definition as a political and legal framework: on the sharing and dissemination of research outputs* | 12/2020 | POLIS N. 19, 2020 http://uet.edu.al/polis/images/1.pdf |
| **V1** | *A policy and legal Open Science framework: a proposal* | 16/09/2020 | https://arxiv.org/abs/2010.04508 https://hal.archives-ouvertes.fr/hal-02962399 https://zenodo.org/record/4075106 |

## Abstract

*It is widely recognised nowadays that there is no single, accepted, unified definition of Open Science, which motivates our proposal of an Open Science definition as a political and legal framework where research outputs are shared and disseminated in order to be rendered visible, accessible, reusable is developed, standing over the concepts enhanced by the Budapest Open Science Initiative (BOAI), and by the Free/Open Source Software (FOSS) and Open data movements.*

*We elaborate this proposal through a detailed analysis of some selected EC policies and laws as well as of the function of research evaluation practices. The legal aspects considered in our examination include, in particular, the study of the role of licenses in the context of the dissemination of research outputs.*

## Motivation

Even if nowadays the Open Access, Free/Open Source Software (FOSS), Open Data, etc. (that we will call Open Science movements from now on) find more and more followers, adepts, and even addicts among the different key actors in the research population, our experience provides everyday examples of scientists that do not know well these movements and their consequences. Some have an idealistic or anecdotal point of view, and many are still not really aware of the deep changes that they do carry on for the research practice.

The use of licenses is well acknowledged in many places, but again, our experience provides examples of scientists that do not even possess a basic understanding of the license issues. Many use software licenses under the advice of colleagues that have no further understanding on licenses, author's rights and their consequences. But the use of a wrong license can have devastating results, opposed to what it was initially expected.

It is also well known within software developer communities that code writers do pay much more attention to the quality of their outcome if it is openly and freely available, and, thus, easily exposed to other developers' examination (and criticisms), which may have real consequences, for example, upon their career. Likewise, researchers that put preprints in places like arXiv[1] do pay special attention to the quality of the initial version in order to attract collaborators, citations, and maybe comments, that will improve the content of the initial work.

Another example of potential consequences of Open Science movements in the research practice appears when noticing the current evolution of the scholarly publishing system. Indeed, there are clear signals that show that the old model dominated by few big (predator) editors is slowly becoming out of date. Yet, part of the research evaluation system is still favouring publications in some Journal titles selected under Impact Factor and Science Citation Index criteria, something that can also become outdated, see for example the analysis in (Guédon, 2001; Delgado-López-Cózar & Martín-Martín 2019; EG to EC, 2019; CODATA CEG, 2020). To illustrate here some of the current publishing evolutions we can mention the recent adoption of an OPEN Roadmap by the

---

[1] https://arxiv.org/

Association for Computing Machinery (ACM) Council as well as the Code Ocean Integrations in the ACM Digital Library to make software and data more discoverable[2].

Despite the increasing presence of Open Science policies and its benefits for the scientific community and the research practices, Open Science can be still considered a young issue requiring, in particular, a deeper understanding of the different ingredients that conform this movement. One of the oldest references on the Open Science subject is (Chubin, 1985), that approaches this concept through the definition of *Closed science*:

*Closed science is defined as research which, in its production, communication, or utilization, is inaccessible to potential consumers. The grounds for such closure are always political, in the sense that certain interests, fortified by legitimate power, can exercise democratic control. The information denied to interested parties becomes the focus of a dispute or controversy which includes the means of control and ways of opening it.*

Paul A. David places in (David, 2008) the *historical formation of key elements in the ethos and organizational structure of publicly funded 'open science'* in the late sixteenth and early seventeenth centuries, where

*... the idea and practice of 'open science' [...] represented a break from the previously dominant ethos of secrecy in the pursuit of Nature's Secrets, to a new set of norms, incentives, and organizational structures that reinforced scientific researchers' commitments to rapid disclosure of new knowledge.*

addressing as well its positive characteristics (David, 2008; p.20):

*... its economic and social efficiency properties in the pursuit of knowledge, and the supportive role played by norms that tend to reinforce cooperative behaviors among scientists.*

Despite these ancient origins, and although Open Science best practices are currently rapidly evolving, real barriers remain for its universal adoption. In particular, many areas of scientific research are still far from being open, as remarked in (Swan, 2012; Morey, Chambers, Etchells, Harris, & Hoekstra, 2019). In our opinion, one important reason for this circumstance has to do with the lack of a clear definition of what Open Science is, which difficults the dissemination of a persuasive message to a wider community of scientists. This deficiency is a well known issue, as you can see for example in the COASP'14 conference of C. Aspesi[3] or as stated in (Tennant, 2018):

*... there is no single, accepted, unified definition or vision of 'open science'...*

This perception is shared by the Organisation for Economic Co-operation and Development (OECD) international organization as it conforms, in (OECD, 2015), the absence of such generally accepted definition, while, at the same time, establishes the key ingredients to be used in the OECD study:

*Open science. There is no formal definition of open science. In this report, the term refers to efforts by researchers, governments, research funding agencies or the scientific community itself*

---

[2] https://www.acm.org/articles/pubs-newsletter/2020/blue-diamond-october2020
[3] http://zeeba.tv/keynote-a-financial-analyst's-perspective-on-open-access/

*to make the primary outputs of publicly funded research results – publications and the research data – publicly accessible in digital format with no or minimal restriction as a means for accelerating research; these efforts are in the interest of enhancing transparency and collaboration, and fostering innovation.*

Contrasting with this imprecise scenery, the Open Science benefits for science and for scientists are widely accepted, as mentioned in (European Commission, 2012), referring to benefits of public investment in research, or as listed in (Tennant, 2018):

*... greater transparency throughout the entire research process, including peer review [...] to combat the 'reproducibility crisis', to expose or prevent research misconduct, to introduce greater accountability for researchers, or to increase the verifiability of the research record in order to engender greater public trust for the scientific enterprise...*

In the same direction, the OECD report (OECD, 2015), the section 4 of (Swan, 2012) and (Fell, 2019) provide further insight about Open Science benefits, including emerging estimations of the economic value of increasing accessibility to public sector research outputs. From a larger, worldwide perspective, the recent CODATA coordinated report (CODATA Coordinated Expert Group, 2020) emphasizes the relevance of Open Science in the starting century:

*Open Science is best characterised as the necessary transformation of scientific practice to adapt to the changes, challenges and opportunities of the 21st century digital era to advance knowledge and to improve our world. This requires changes in scientific culture, methodologies, institutions and infrastructures.*

Note that this report also studies in its section 8 some of the negative impacts of Open Science and how to address them.

Furthermore, in this 2020 pandemic year, there is little doubt in the need of a sound, extended Open Science, and that its benefits spread to the whole society, worldwide (Arrizabalaga, Otaegui, Vergara, Arrizabalaga, & Méndez, 2020; CODATA Coordinated Expert Group, 2020; EC DGRI, 2020; UNESCO, 2020)[4].

Getting back to our main issue here, that is, on the need for a clear and broad definition, let us observe that the authors of (Vicente-Saez & Martinez-Fuentes, 2018) explicitly acknowledge this fact:

*... there is a lack of awareness about what Open Science is, mainly due to the fact that there is no formal definition of Open Science ...*

after performing a thorough review of publications of the period 1985-2016, ending up by expressing the need for further research to clarify this concept.

It is our aim, in the present work, to contribute towards the fulfilment of this purpose.

---

[4] See also the UNESCO video at YouTube: Open Talks Webinar "Open Science for Building Resilience in the Face of COVID-19" https://www.youtube.com/watch?v=nbwjEZ1n1Gg

Rather than to concentrate here in the philosophical aspects involved in the definition of Open Science (Fecher & Friesike, 2014), we will focus in more practical endeavours, represented by the sharing and dissemination conditions of the research production, or by *the norms and rules governing disclosure* of these research outputs (in the words of Paul A. David), that is *norms of full disclosure and cooperation in the search for knowledge* (David, 2008; p. 23).

Thus, in the following section we will present the three fundamental components supporting our Open Science definition proposal as *a political and legal framework where research outputs are shared and disseminated in order to be rendered visible, accessible, reusable*. Section 3 emphasizes the relevant role of licenses in this context. Section 4 describes, through some examples, our proposal for an Open Science framework comprising policies, laws and research evaluation practices. This work ends by describing the advantages of our proposal, which are argued in the final section and exemplified via the analysis of a recent case covered by the media.

## Three pillars for an Open Science vision

There is an extended literature studying Open Science key issues and, although it is a pending task, it is not the goal of our present work to provide a bibliographic survey on the foundations of Open Science. Rather, in this section, we summarily present what we consider the three pillars where our Open Science vision stands over.

We place our first pillar on the BOAI, the Budapest Open Access Initiative (2002)[5], that defines *open access to the scholarly journal literature* as follows:

*By "open access" to this literature, we mean its free availability on the public internet, permitting any users to read, download, copy, distribute, print, search, or link to the full texts of these articles, crawl them for indexing, pass them as data to software, or use them for any other lawful purpose, without financial, legal, or technical barriers other than those inseparable from gaining access to the internet itself. The only constraint on reproduction and distribution, and the only role for copyright in this domain, should be to give authors control over the integrity of their work and the right to be properly acknowledged and cited.*

Let us remark that a historic vision, initiated by Peter Suber, of the Open Access movement can be found at the Open Access directory (OAD)[6], with a pioneer online library dated back to 1966.

The Open Access movement launched by the BOAI was partially inspired by the Free/Libre/Open Source Software movements (FOSS or FLOSS[7]), which constitute, for us, the second pillar of our Open Science vision. Free software was defined[8] by the Free Software Foundation (FSF), launched by R. M. Stallman (1985), as follows:

*"Free software" means software that respects users' freedom and community. [...]*
*The four essential freedoms. A program is free software if the program's users have the four*

---

[5] https://www.budapestopenaccessinitiative.org/
[6] http://oad.simmons.edu/oadwiki/Timeline
[7] FLOSS stands for Free/Libre/Open Source Software, where the French/Spanish word *Libre* enhances the freedom philosophy of the Free Software movement.
[8] See https://www.gnu.org/philosophy/free-sw.en.html. See also The Open Source Initiative web page http://www.opensource.org/docs/osd for the definition of Open Source Software (OSS).

*essential freedoms:*

- *The freedom to run the program as you wish, for any purpose (freedom 0).*
- *The freedom to study how the program works, and change it so it does your computing as you wish (freedom 1). Access to the source code is a precondition for this.*
- *The freedom to redistribute copies so you can help others (freedom 2).*
- *The freedom to distribute copies of your modified versions to others (freedom 3). By doing this you can give the whole community a chance to benefit from your changes. Access to the source code is a precondition for this.*

The third pillar is based in the Open Data movement. As we are not aware of older initiatives, we place its initial step in CODATA, the Committee on Data for Sciences and Technology launched by the International Council of Scientific Unions (ICSU) in 1966 (CODATA, 1971; p. 2):

*CODATA est un Comité au niveau scientifique international le plus élevé [...]*
*à cause de l'importance qui s'attache à l'évaluation des données [...]*
*c'est un comité de coordination et sa principale tâche est de prendre des initiatives et de souligner l'importance des aspects communs à plusieurs domaines de la science et de la technologie, ce qui comprend les activités suivantes:*
*a) l'évaluation des méthodes de contrôle de la qualité,*
*b) la définition des besoins des utilisateurs,*
*c) les standards divers,*
*d) les techniques de l'information [...]*

*[CODATA is a Committee at the highest international scientific level [...]*
*because of the importance attached to the evaluation of data [...]*
*it is a coordinating committee and its main task is to take initiatives and highlight the importance of aspects common to several fields of science and technology, which includes the following activities:*
*a) evaluation of quality control methods,*
*b) the definition of user needs,*
*c) various standards,*
*d) information technology [...][9]]*

As we can see in (CODATA, 1971; p. 3), data issues were already part, by that time, of the international scientific considerations:

*The compilation of evaluated numerical and other quantitative scientific data is an important part of the general problem of a Science Information System which encompasses abstracting, storage and retrieval of unevaluated scientific information as well as the evaluation of this information in the form of selected and critical sets of quantitative data, including critical review papers.*

Moreover, in (CODATA, 1972), we can find *A guide to procedures for the publication of thermodynamic data,* which shows how data dissemination issues were, even then, at the heart of the CODATA considerations.

---

[9] The authors provide their own translation to French citations. Authors prefer to keep the original text for French readers to enjoy it, very much in line with the Helsinki Initiative on Multilingualism in Scholarly Communication (2019), see https://doi.org/10.6084/m9.figshare.7887059.

A modern version of this kind of international initiative to deal with scientific data issues is the Research Data Alliance (RDA)*[10]*, launched as a community-driven initiative in 2013 by the European Commission, the United States Government's National Science Foundation and National Institute of Standards and Technology, and the Australian Government's Department of Innovation. RDA's goal is to build the social and technical infrastructure to enable open sharing and re-use of data.

Besides these three key points that we have swiftly described, there are other scientific initiatives and movements that do enrich our Open Science vision. We would like to mention here the Reproducible Research initiative[11]:

*An article about computational science in a scientific publication is not the scholarship itself, it is merely advertising of the scholarship. The actual scholarship is the complete software development environment and the complete set of instructions which generated the figures. (D. Donoho)*

as well as the *open peer review* studies and the current evolutions in its practices (Ross-Hellauer, 2017; Morey et al., 2019; Tennant, 2018). See for example (UNESCO, 2020) for a more complete vision.

## Legal aspects: the important role of licenses

Our three Open Science pillars set the dissemination conditions for research outputs, where licenses play an important role. A software is free if it is released complying with the four freedoms of the above mentioned definition given by the FSF; and this compliance should be stated in the license, a (legal) document that is included in the set of files that constitute the software, and that comprises the source code, the compiled code, documentation, etc. The running, loading, reproducing, translating or arranging of a computer program can only be done upon the corresponding (written) authorisation (Directive 2009/24/EC, 2009), and, if there is not license, then "*All rights are reserved*". In particular no one outside the circle of the authors and the rightholders can run the software (legally speaking). A license sets then *the sharing conditions* of the disseminated software.

Thus, another important point of the Free Software Foundation was to develop the GNU GPL license, a license according to its philosophy and that accompanies its software production[12].

This is a curious difference between the Open Access movement and the FOSS movement, as it took some time in the Open Access movement to speak about the use of licenses in the dissemination of written works, mainly mentioning the Creative Common (CC) licenses[13]. For example section 6.2 of (Swan, 2012) indicates that

*… if an article carries no licence information at all it is not clear to users what they might do with it […] Proper, appropriate licensing sets out the conditions for re-use and reassures would-be users that they can use the material in particular ways with impunity.*

---

[10] https://www.rd-alliance.org/about-rda  
[11] http://reproducibleresearch.net/  
[12] See the History of the GNU GPL at https://www.free-soft.org/gpl_history/. For further information on the GPL license and its consequences on the evolutions of the information technology market see Eben Moglen's plea for Free Software before the European Parliament (2013-07-09) at https://www.youtube.com/watch?v=FI1CoeqyD5o  
[13] https://creativecommons.org/licenses/

Once you have access to a document, there is not legal barrier to its reading. So the important point for the Open Access movement was to have access to the scientific literature. On the other hand, even if you have access to a software, its use can encounter legal barriers. This is why the first freedom (freedom 0) of the free software definition is about the use of the software. Furthermore, freedom 1 and freedom 3 are related to the access to the source code in order to be able to study it and to produce new software. These principles correspond to the production of (new) scientific knowledge, which should be eased by the open access to the existing scientific literature.

Nowadays, despite these initial differences, the important role of licenses is now clear for the Open Access movement, as we can see in recent publications such as (Arrizabalaga, et al., 2020; EC DGRI, 2020; UNESCO, 2020)[14]. Licenses are in fact the tool to set aside the legal barriers referred to in the BOAI open access definition.

Other important issues related to licenses that we introduce here deal with:
*i)* the relation between the open access definition and the most common used licenses (such as CC),
*ii)* the adequacy of the use of CC licenses for software, documents and data,
*iii)* the way to overcome the possible limitations imposed by a license.

Regarding item *i)*, note that the dissemination of research outputs such as articles, in compliance with the BOAI open access definition, can be realized with the CC licenses CC-BY and CC-BY-SA, but not with the others (CC BY-ND, CC BY-NC, CC BY-NC-SA, CC BY-NC-ND), as there are restrictions in the production of derivative works (as for example the translation of a document) or in the commercial use of the outputs. It is also possible to use Public Domain marks[15] (Public Domain, CC-0) and they also comply with the BOAI open access definition.

Corresponding to item *ii)*, we remark that CC licenses are not adapted to software dissemination[16]. Software licenses can be found, for example, in the Free Software Foundation (FSF) web site[17], in the Open Source Initiative (OSI) web site[18] or in the Software Package Data Exchange (SPDX) web site[19].

On the other hand, we observe that the Version 4.0 of the Creative Commons (CC) licenses was published in 2013 and was developed to be more user-friendly and more internationally robust, and to cover more explicitly *sui generis* database rights[20] (Directive 96/9/EC, 1996; Labastida & Margoni, 2020) so they can also be used in the dissemination of databases.

Finally, let us remark that, regarding item *iii)*, each license sets its own sharing (legal) framework giving rights but also conditions that should be respected. In the case that the planned use of the research output does not agree with the default conditions given by the license, it may be possible to contact the rightholders to set other agreements or collaboration contexts, that is, *other sharing conditions*.

---

[14] See also the Webinar "*Abrir con Propósito en América Latina. Una reflexión de como construir equidad e inclusión estructurales*" organized by the UNESCO and other Organizations the 23 October 2020, see the announcement at http://amelica.org/index.php/2020/10/07/abrir-con-proposito-en-america-latina-una-reflexion-en-como-construir-equidad-e-inclusion-estructurales/ and the video at https://youtu.be/l9aC_sw7Xtc. In particular the presentation of Eduardo Aguado López insists on the role of licenses.

[15] https://creativecommons.org/share-your-work/public-domain/

[16] https://creativecommons.org/about/program-areas/software/

[17] https://www.gnu.org/licenses/

[18] https://opensource.org/licenses

[19] https://spdx.org/licenses/

[20] See https://creativecommons.org/version4/ and https://wiki.creativecommons.org/wiki/4.0.

## An Open Science framework: policies, laws and research evaluation practices

We consider that a proposal for a sound Open Science framework has to declare, at least, its position concerning two basic issues: the related political and legal aspects. In this section we describe our perspective on these core points, highlighting as well the consequences of our choices on a practical context that involves research evaluation practices.

Our proposal for a sound Open Science framework begins with the adoption of Open access policies such as the ones decided by the European Commission (EC) as an answer to the Open Access movement and that has have deep consequences in the EC research founding program Horizon 2020 (H2020). In its communication *"Towards better access to scientific information: Boosting the benefits of public investments in research"* (European Commission, 2012), the Commission:

*...sets out the action that intends to improve access to scientific information and to boost the benefits of public investment in research. It also explains how open access policies will be implemented under 'Horizon 2020', the EU's Framework Programme for Research and Innovation (2014-2020). [...]*
*To improve access to scientific information, Member States, research funding bodies, researchers, scientific publishers, Universities and their libraries, innovative industries, and society at large need to work together [...] so that the 'fifth freedom' of the EU – the free circulation of knowledge – can become a reality.*

These policies go along with guidelines to explain the rules on open access to scientific peer reviewed publications and research data that beneficiaries have to follow in projects funded or co-funded under Horizon 2020 (EC DGRI, 2017). Among other actions and EC funded projects we can find the OpenAIRE project[21] with the mission to[22]

*Shift scholarly communication towards openness and transparency and facilitate innovative ways to communicate and monitor research.*

To fulfill this mission, OpenAIRE provides Open Science services and participate *to foster the open science dialogue for policies and their implementation in Europe and beyond,* see the "Open Science overview in Europe" page in the OpenAIRE web site[23] to get more information about Open Science policies in Europe.

An Open Science policy renders publicly financed research outputs open. But it is understood that there may be some hinders, and, if it is the case, researchers should explain the reasons to keep the outputs closed, at least for some period. The European Commission mantra is a*s open as possible, as closed as necessary* (EC DGRI, 2017).

---

[21] https://www.openaire.eu/
[22] https://www.openaire.eu/mission-and-vision
[23] https://www.openaire.eu/os-eu-countries

The second part of our proposal recognises the need to turn such policies into legal provisions, as has been done in the French 2016 law for a Digital Republic Act (Loi n. 2016-1321 pour une République numérique, 2016), by setting up the legal framework for open access to scholarly communication[24] and creating thereby a new right for researchers, where authors can give open access to copies of their publications *even if they have granted the copyright to a publisher*. This law also sets the legal framework for open data and goes along with a *Décret* (Décret n. 2017-638, 2017) to list the licenses that should be used for data and software[25]. Other licenses are possible, but there is an approval process to be applied[26].

This French law has been followed by a policy document, the *National plan for Open Science* published by the *Ministère de l'Enseignement supérieur, de la Recherche et de l'Innovation (MESRI, 2018).* This plan contains three main axes dedicated to publication's open access, research open data and a sustainable, European and international dynamic framework. Each axis establishes three measures, including:

*1. Make open access mandatory when publishing articles and books resulting from government-funded calls for projects. [...]*
*4. Make open access dissemination mandatory for research data resulting from government-funded projects.*

These two components (policies, legal) of our proposal are direct consequences of the emphasis of the Open Science movements on these issues. But still there is a long path to be built in order to make Open Science practices to become part of the everyday practices in research.

In our view, there are three keystones that have to be considered to pave this path:
- the required evolution of policies of Universities and research performing organizations,
- the development of Open Science-oriented infrastructures and services,
- the transformation of evaluation policies and practices.

We have already mentioned in this section the EC evolutions in Open Science policies and the consequent evolutions in national laws and policies. Another example of Open Science policy evolution at large scale, perhaps one of the most relevant at the time of writing this work, is the ongoing UNESCO global consultation initiative on Open Science[27]. Let us recall that

*UNESCO's mission is to contribute to the building of peace, the eradication of poverty, sustainable development and intercultural dialogue through education, the sciences, culture, communication and information...*

and that one of the purposes and functions of this Organisation is to *maintain, increase and diffuse knowledge* (Swan, 2012). The UNESCO Open Science global consultation aims to build a UNESCO Recommendation on Open Science, following the commission of the 193 Member

---

[24] The second chaper entitled *Économie du savoir* refers to *écrit scientifique issu d'une activité de recherche financée au moins pour moitié par des dotations de l'Etat*, and, as a consequence, also includes software.

[25] Note that, in here, software and data refer to the French law concept that can be larger and include the research outputs like software and data produced by research teams.

[26] Please note that, the list of licenses available at https://www.data.gouv.fr/fr/licences does not include currently the Creative Common licenses nor the European Union Public Licence (EUPL) for software, available at https://joinup.ec.europa.eu/collection/eupl/eupl-text-eupl-12.

[27] https://en.unesco.org/science-sustainable-future/open-science/consultation

States at the 40th session of UNESCO's General Conference in order to develop an international standard-setting instrument on Open Science in the form of a UNESCO Recommendation on Open Science (UNESCO, 2020; CODATA Coordinated Expert Group, 2020). As a consequence, Universities, funders and research organizations are reviewing their policies in order to adapt to law and policy changes and to adopt and implement an answer to Open Science requirements.

To illustrate more in detail this first keystone, we would like to mention some examples of such evolutions. The first example corresponds to the local environment of one of the authors of this work. The newly named Université Gustave Eiffel[28] is the result of the fusion of several academic organizations including the Université de Paris-Est Marne-la-Vallée (that hosts the Laboratoire d'informatique Gaspard-Monge (LIGM)) and the Institut Français des Sciences et Technologies des Transports, de l'Aménagement et des Réseaux (IFSTTAR). This University has been launched on January 1st 2020, and one of the first decisions of the Research Vice presidency has been to start a working group in order to elaborate the Open Science policies of the University, the *Groupe de Travail "Politiques recherche ouverte à la société" (GTPROS)*. This group counts with several subgroups, with two of them dedicated to data and software, counting with the participation of one of the authors of this paper.

Other examples to show University Open Science evolutions correspond to the Göttingen eResearch Alliance and the central role of the Vilnius University Library. The Göttingen University institutional data management policy was published in 2014, and then the eResearch Alliance was established to serve research projects and provide direct support to researchers, combining library and IT services and expertise (Schmidt & Dierkes, 2015). The Vilnius University Senate approved the Regulations of Open Access to the University Scientific Works and the Results of Scientific Research prepared by the Vilnius University Library in 2009. The Vilnius University Library has then the task to develop the scholarly communication tool dedicated to sustain open access to information and open science (Kuprienė & Petrauskienė, 2018).

Both articles highlight the great value of the direct collaboration with researchers, like, for example, embedding research-data managers into research teams (Göttingen), or their participation into determine roadmaps and priorities for the infrastructures and services under development (Vilnius).

Both examples show as well how the evolution in policies goes along with the development of infrastructures and services to accompany researchers with their output's dissemination, as there is still a lot of work to be done – the second keystone – in order to tackle the current imbalance between already imposed legal requirements for researchers, and the still ongoing and unfinished work to build the necessary infrastructures and services.

The human resources in charge of developing the needed infrastructures and services are also simultaneously and increasingly acquiring the necessary knowledge and expertise to deal with these challenges. The institution decision-makers are in a similar evolving situation. As a consequence, researchers have to adapt (and sometimes to improvise with) their working mechanisms to the new policies and funding requirements, while contributing to their implementation by the expression of their new, urgent research needs.

---

[28] https://www.univ-gustave-eiffel.fr/

In this context, we can also mention the IFSTTAR experience of an institutional data deposit[29] developed to answer the researchers' needs. Its development towards building the institutional data deposit for the Gustave Eiffel University is currently under consideration at the GTPROS.

At European level, the European Open Science Cloud (EOSC) is the Open Science infrastructure currently under construction, see for example (HLEG EOSC, 2016; Budroni, Burgelman, & Schouppe, 2019).

But the keystone that we analyse here in more detail is the necessary evolution in research evaluation practices, as it has been pointed out in (ALLEA, 2012; EG to EC, 2019), with procedures like the ones proposed in (McAllister, Esposito, O'Carroll, Vandevelde, Maas, et al., 2017; Gomez-Diaz & Recio, 2019). A well known example of open access evaluation evolutions is provided by the University of Liège (Belgium) and the policy evolutions introduced by its Rector, Professor Bernard Rentier, in 2007 (Swan, 2012).

Research evaluation happens in recruitment and career progression, and the institution's evaluation policies will drive the selection of best candidates with Open Science good practices. It is also a general practice of research institutions and funders to evaluate the laboratories, institutes, research units... regularly, and evaluation policies and practices should evolve from now on to include best Open Science practices.

Research evaluation also happens in project funding: the funder policies do establish under which conditions the projects are selected for funding, as well as to give the instructions for free/open access of the project outputs. The final evaluation of the project assesses the quality of the submitted work and verifies if the outputs are open in compliance with the open access instructions. In the case in which the outputs are not open, the evaluators should consider the alleged reasons for the avoidance of public access, but, in any case, they need to have access to the outputs in order to realize a sound evaluation.

Other evaluation contexts appear, for example, when looking for collaborators for a project, or for a publication, or in the selection of a journal for a publication. This is what we call here the *"research community evaluation"*, in which the perception of a colleague (or of a journal) reputation plays an important role. For example, authors or reviewers can choose or avoid a journal following its open peer review practices. They can also stand for more "openness" when they are contacted for their participation in a new project. This is the power of the research community and the challenge of the *reputation system*.

Funder and institution evaluations and research community evaluations are therefore a powerful tool to enhance effective Open Science evolutions, and constitute, in our view, the third cornerstone in the Open Science path. But, *as the cat biting its own tail,* the evaluation wave can only play fully its role if policies and laws are well into place.

In summary, without the necessary evolution in policies and laws, without a sound research evaluation system to enhance full compliance with the resulting policies and laws, it is not possible to achieve a solid fulfillness of the Open Science movements.

---

[29] https://research-data.ifsttar.fr/

## A digression: the The Lancet case

*On 30 January 2020 the World Health Organization (WHO) declared that the SARS-CoV-2 outbreak constitutes a Public Health Emergency of International Concern (PHEIC)* (EC DGRI, 2020). As a consequence, the European Commission stressed the urgency *to provide immediate open access to their related publications, data and any other output possible* for researchers receiving H2020 funding in SARS-CoV-2 related research, following the guidelines of (EC DGRI, 2020). In particular, the OpenAIRE project centralized in its web site[30] the information comming from the Zenodo Community, from the OpenAIRE Gateway, the RDA COVID-19 Fast Track Working Group and from other EC ongoing efforts. Moreover, researchers, funders and some publishers expressed the importance of open access to research outcomes in this context, as we can see for example in (Arrizabalaga et al., 2020; CODATA Coordinated Expert Group, 2020; EC DGRI, 2020; UNESCO, 2020).

But the urgency of any situation like this one should remain, more than ever, incompatible with speedy publications of poor scientific content. In the current pandemic context, many papers have been delivered in preprint servers without proper peer review, and some high-profile journals have published papers that have been retracted (CODATA Coordinated Expert Group, 2020). Moreover, as remarked in (Arrizabalaga et al., 2020), some of the open access publications available on the publishers platforms have not license at all, and thus the publisher can revoke this kind of access at any time, and then, further study can be restricted in the future.

Among these publications, possibly the most unhappily famous is the one of the The Lancet journal, that has been fully retracted[31]. This publication deals with the use of the Hydroxychloroquine drug for the treatment of Covid-19 disease, as presented for example in the French media[32]. Let us recall that all The Lancet journals endorse the Wellcome Trust Statement[33] to ensure that relevant research findings and data are shared rapidly and openly.

We will no enter here in the scientific details of this work, as they can be found for example in this entry blog[34] of the Barcelona Institute for Global Health (ISGlobal)[35]. Rather we consider here some of the involved Open Science aspects.

In general, data gathered and used to produce a publication should be available for co-authors and for reviewers before publication, but maybe the data has ethical or personal issues, and then it should be available for a restricted number of persons in a restricted environment for its study in order to allow the validation of the work. Reviewers should be given the correct time to referee a scientific work. Yet, the Covid-19 urgency has restrained the reviewing period to four days. But in the case this interval is considered too short for doing a correct assessment, reviewers should ask for more time or retract from the referee process. And this has not happen in the The Lancet case.

---

[30] https://www.openaire.eu/openaire-activities-for-covid-19

[31] https://www.thelancet.com/journals/lancet/article/PIIS0140-6736(20)31180-6/fulltext

[32] FranceInfo, *Vrai ou Fake section*, Coronavirus : le "LancetGate" ou le naufrage de la science business, 20-06-2020 https://www.francetvinfo.fr/sante/maladie/coronavirus/video-coronavirus-le-lancetgate-ou-le-naufrage-de-la-science-business_4014381.html.

[33] https://wellcome.ac.uk/coronavirus-covid-19/open-data

[34] https://www.isglobal.org/en/healthisglobal/-/custom-blog-portlet/sin-rigor-y-transparencia-no-hay-ciencia-sobre-surgisphere-y-sus-publicaciones-cientificas-en-revistas-de-alto-impacto/93337/0

[35] https://www.isglobal.org/en/

This affair shows the importance of having open access to publications, which has allowed, in this case, the rapid post-publication evaluation and a fast reaction of the scientific community, yielding the detection of several irregularities of this work and, thus, its retraction. It also shows the way to go forward in order to react after suffering these kind of problems. Indeed, authors who have been involved in such situations are prone to experience further issues when seeking for new collaborations or for new project funding, unless their research practices and accountability methods evolve to adopt and follow clear and transparent Open Science rules. Similar implications can be formulated concerning journals, as they can be also exposed to negative "research community evaluation", being prompted then to adopt more sound Open Science best practices, like open peer review.

## Discussion

We have initiated this article reflecting on the lack of a clear definition of Open Science, and on the need to contribute towards its clarification. As supported by the ideas developed in this work, it is our vision that Open Science is the framework that renders research outputs visible, accessible, reusable. Thus, it is most important to analyse what outputs are disseminated and by who, and when. Next, once the dissemination conditions have been finally decided, it is crucial to establish – in order to render the outputs visible, accessible, reusable – how the outputs are disseminated, in which sharing conditions, and in which places researchers will do the dissemination.

As we have introduced in section 3 and placed into context in section 4, the sharing conditions of the research outputs do involve scientific (political) movements, policies, and laws and legal issues (author rights, licenses). It involves also the producer's choices (Gomez-Diaz, 2015) (when and where outputs are made accessible, and in which conditions they can be reused), and the places where the dissemination is realized, that can include Journals (scientific journals publishing articles, data papers and/or software papers, etc.) and other infrastructures and services for preprints, data, software and other research outputs deposit, output search and retrieval interfaces, and that do provide or facilitate the outputs' reuse. Therefore, we conclude that Open Science should refer to *the political and legal framework where research outputs are shared and disseminated in order to be rendered visible, accessible, reusable*.

Landscaping is evolving quickly nowadays, as policies and laws evolve at many levels, and it is becoming more common to have free/open access to scientific publications as well as to other research outputs like software and data. As mentioned in section 4, research evaluation is the tool to reinforce the basis of the Open Science political and legal framework, enabling it to reach its final goals, and, thus, spreading its benefits everywhere.

Evaluation happens every day at every level, from the selection of collaborators to the recruitment and career evolutions, including as well peer reviewed publications and research funding decisions. Research evaluation is not the only keystone to enhance Open Science implementation, there is also a need to develop the infrastructures in which the outputs are accessible and reusable, as well as the supporting services to accompany the researchers. These items (research evaluation practices, infrastructures and services) are evolving to adapt to this, somehow new, Open Science context, both at supra-national level (e.g. UNESCO, EOSC) and at national and local levels which include the evolutions of Universities and other research performing institutions.

Would have been more established this Open Science framework, the The Lancet retracted publication would have neither been co-authored[36], nor passed the peer-reviewed stage, or been accepted by editors for publication.

## Conclusion

In this paper we have introduced some examples of the pros and cons of the current progression of the Open Science movement. Reacting to the declared need to advance toward a more clear understanding of the Open Science concept, we have highlighted three relevant initiatives (BOAI, FOSS, Open data movements) that conform our Open Science vision, emphasizing their role on the dissemination procedures for research outputs, and supporting our vision of Open Science as *the political and legal framework where research outputs are shared and disseminated in order to be rendered visible, accessible, reusable*.

The important role of licenses in this context has been reported in section 3 as a key ingredient to understand our proposal for a sound Open Science framework that includes EC promoted open access policies and legal provisions such as the French Digital Republic Act. We have as well developed the needed elements to put in to practice the proposed framework, placing the accent over its implementation with research evaluation practices.

We have concluded bringing up some suggestions for policies, laws and evaluation systems to evolve in order to achieve the goals of Open Science.

**Acknowledgements**

This work benefits of the interesting and fruitful exchanges at the GTPROS working group of the University Gustave Eiffel. It is also partially funded by the *Vice-présidence international* of this University and by the *Laboratoire d'informatique Gaspard-Monge*.

Versions of this work are available at arXiv (https://arxiv.org/), HAL (https://hal.archives-ouvertes.fr/), POLIS (http://uet.edu.al/polis/) and Zenodo (https://zenodo.org/), under the Creative Commons Attribution 4.0 International License (CC-BY) (http://creativecommons.org/licenses/by/4.0/).

---

[36] Due to the lack of access to the research data by some of the authors, as reported in the news.